# An improved continuous compositional-spread technique based on pulsed-laser deposition and applicable to large substrate areas

H.M. Christen, C.M. Rouleau, I. Ohkubo, H.Y. Zhai, H.N. Lee, S. Sathyamurthy, and D.H. Lowndes

*Oak Ridge National Laboratory, Condensed Matter Sciences Division, Oak Ridge, Tennessee 37831-6056*

**Abstract**

A new method for continuous compositional-spread (CCS) thin-film fabrication based on pulsed-laser deposition (PLD) is introduced. This approach is based on a translation of the substrate heater and the synchronized firing of the excimer laser, with the deposition occurring through a slit-shaped aperture. Alloying is achieved during film growth (possible at elevated temperature) by the repeated sequential deposition of sub-monolayer amounts. Our approach overcomes serious shortcomings in previous *in-situ* implementations of CCS based on sputtering or PLD, in particular the variations of thickness across the compositional spread and the differing deposition energetics as function of position. While moving-shutter techniques are appropriate for PLD-approaches yielding complete spreads on small substrates (i.e. small as compared to distances over which the deposition parameters in PLD vary, typically ≈ 1 cm), our method can be used to fabricate samples that are large enough for individual compositions to be analyzed by conventional techniques, including temperature-dependent measurements of resistivity and dielectric and magnetic and properties (i.e. SQUID magnetometry). Initial results are shown for spreads of $(Sr_{1-x}Ca_x)RuO_3$.

**Introduction**

Recent advances in data analysis, data acquisition, and robotic techniques have lead to intensive efforts to develop combinatorial techniques, i.e. approaches in which a multitude of samples are prepared in one fabrication run. The clear advantage over the conventional "single-sample approach" is that all fabrication parameters are kept identical by design, while the effect of chosen variables (such as composition, film deposition temperature, or film thickness) can be studied in much detail. Discrete combinatorial techniques have been very successful in pharmaceutical research, and a corresponding precursor-based thin-film approach has been developed for various inorganic materials that can be synthesized via an equilibrium route.[1] Here, the starting materials are first deposited at room temperature, and a subsequent annealing process can lead to intermixing and crystallization. The same precursor principle can also be applied to continuous compositional-spread (CCS) approaches, where the composition varies uniformly and continuously across the sample.[2-4] However, for materials that are not formed by the equilibrium processes involved in the precursor method, and in general for materials with properties that depend strongly on the deposition technique, alternate approaches must be developed that are based on the *in-situ* formation of the desired alloys.

A CCS approach providing a means for this *in-situ* alloy formation has been introduced more than 35 years ago based on the co-deposition of the constituents: Kennedy[5] showed that meta-stable intermetallic phases can be formed by co-sputtering from multiple sources, and that a substrate holder with a spatial temperature gradient can be used to determine the optimal deposition temperature for each composition. The approach was later refined by Hanak,[6] but did not gain popularity before the widespread use of electronic data analysis and representation. The successful application to high-k dielectrics[7] has lead to an increased interest in sputter-based CCS, with recent work focusing, for example, on transparent conducting oxides.[8]

Pulsed Laser Deposition (PLD) can similarly be used for a co-deposition based CCS approach. In particular, a beam splitter may be used to obtain simultaneous ablation from two PLD targets, and the spatially overlapping plumes result in a composition variation in the deposited film.[9] As an alternative to co-deposition, *in-situ* intermixing between the constituents of an alloy can also be achieved by the repeated, sequential deposition of sub-monolayer amounts of the individual constituents, as demonstrated by the successful use of rotating, segmented targets to form uniform films of mixed oxides.[10] Applying this approach to CCS is easily achieved by using a synchronized target and substrate positioning approach, so that the plumes of the individual constituents are spatially centered on different positions on the substrates. As we have previously shown, this yields a CCS approach that can be used for the growth of meta-stable oxides[11] and epitaxial superlattices.[12]

One advantage of these *in-situ* approaches is that no masks are required, which simplifies the process and reduces the risk of cross-contamination. Unfortunately, such PLD and sputter-based methods also suffer from serious drawbacks. In particular, the film thickness varies as function of position on the substrate, and thus the comparison of different compositions is convoluted with that of varying film thickness. More importantly, the deposition energetics are also different for each composition, which is a significant fundamental problem particularly for the PLD-based approaches.

Furthermore, the composition is a non-trivial function of the position that needs to be determined from trial runs and calibration procedures.

These difficulties can be overcome by introducing a moving mask between the substrate and the target.[13] Again, the deposition sequence is chosen such that less than a monolayer of each constituent is deposited at one time (for each "pass" of the mask), assuring complete intermixing, i.e. *in-situ* alloy formation. To guarantee that the film thickness and the deposition energetics are reasonably uniform across the entire composition spread, the sample size should be less than about 1 cm. These dimensions are appropriate for specialized characterization tools, such as optical characterization for phosphor materials, concurrent x-ray diffraction,[13] SQUID microscopy,[13] or microwave microscopy.[14] Other characterization techniques, including those most commonly used in solid-state physics, require larger sample sizes. For example, careful four-circle x-ray analysis, ellipsometry, and temperature-dependent measurements of resistivity, dielectric permittivities, and magnetic properties (using a SQUID magnetometer) all require sample areas on the order of 10 mm$^2$ per measurement point, and thus a much larger area for a compositional-spread.

In this work, we introduce a new approach to CCS in which the composition variations occur over a range of several centimeters, thus allowing for these measurements to be performed for various compositions. For the exploration of binary phase diagrams, the only fundamental limit on the sample size is the travel range of a translating substrate heater.

In this paper, we demonstrate the approach by showing results for composition-spread films of $(Sr_{1-x}Ca_x)RuO_3$. This particular alloy (with complete solubility for $0 \leq x \leq 1$) is chosen for its interesting magnetic properties[15]: Ferromagnetism is observed at low Ca concentration $x$ (with a $T_c$ of about 160 K for SrRuO$_3$ and decreasing with increasing $x$), whereas the susceptibility of CaRuO$_3$ shows a negative Weiss temperature, suggesting a tendency toward antiferromagnetism. In addition, thin films have been grown previously by various groups, allowing us to compare the quality of the films obtained here to literature values.

The paper is structured as follows: First, we introduce the basic principle of the deposition and present the key elements of the control algorithm. We then show that the desired profiles can be obtained accurately. Next, the approach is applied to binary phase diagrams of (Sr$_{1-x}$Ca$_x$)RuO$_3$ for which we show uniform composition variations across the sample. Finally, we show that the obtained films are of good quality, and that measurements of lattice parameters (from x-ray diffraction), resistivity (i.e. $\rho(T)$ curves) and magnetic properties (i.e. M(H) loops) can be obtained.

**Method**

Our PLD-CCS approach is based on a conventional PLD system, modified only by having full and rapid control over target exchange, heater translation, and laser firing. A Lambda Physik LPX325i excimer laser with 248 nm radiation (160 W) is used to ablate from any one of four targets mounted on a standard carrousel. The substrates are attached (using silver paint) to a rotatable Inconel plate, which is heated radiatively from behind by an exposed metallic filament. The entire heater/substrate assembly is mounted on a translation stage capable of 75 mm of travel. Appropriate stepper motors assure that any position along that line can be reached in less than 0.5 s. Target-substrate distance is typically 50 – 80 mm, and a slit-shaped aperture (typically a 3 mm wide opening) is inserted 3 mm from the substrate surface.

For the deposition of composition spreads, the substrate is passed behind the slit-shaped aperture and the laser is fired at pre-defined "trigger points". These non-equally spaced points are selected such as to lead to the desired spatial variation of the composition, as described below. After one pass, during which less than one monolayer of the material must be deposited, the target is exchanged, a different set of trigger points for the second material is chosen, and the substrate is passed behind the aperture again. Figure 1 illustrates this procedure schematically.

In order to obtain a composition that varies linearly from left to right across the substrate, both constituents have to be deposited with linear spatial deposition-rate profiles. In other words, the deposition of each constituent must be such that if deposited by itself, the resulting film would have a "wedge"-type thickness profile ($t(x) = t_0 + ax$). Obtaining a linear profile via a slit-shaped aperture is significantly different from using a moving-edge mask (allowing deposition onto a variable fraction of the substrate). With moving-edge masks, linear wedges are obtained simply by a constant-velocity motion and a constant deposition rate, but the approach is inadequate for large substrate areas.

In the current PLD-approach, each laser pulse leads to the deposition of a finite and equal amount of material in the narrow region of the aperture slit. This deposition can be described as a smudged line perpendicular to the direction of the composition gradient, with a cross-section that roughly follows a Gaussian shape. We therefore need to approximate the linear profile as a superposition of $N$ Gaussians centered at $x_i$ (these values are then used as "trigger points" during the deposition).

Simple geometrical arguments show that a linear profile $t(x) = t_0 + ax$ can be approximated by such a superposition if successive trigger points are spaced by

$$\delta x_i = -\sum_{n=1}^{i-1} \delta x_n - \frac{t_0}{a} + \sqrt{\left[\sum_{n=1}^{i-1} \delta x_n + \frac{t_0}{a}\right]^2 + 2\delta t_1 \frac{t_0}{a} + \delta x_1^2} \,. \tag{1}$$

$\delta x_0$ is chosen iteratively such that the desired length of the profile corresponds to

$$L = x_N + \frac{\delta x_N}{2}. \tag{2}$$

The calculations in Fig. 2 show the type of profile that would be obtained if 15 trigger points are used and the deposition is assumed to have a Gaussian shape with a width of 0.1. The desired profile for material A is shown as broken line in Fig. 2a and can be represented as

$$t(x) = \begin{cases} 0.8 & x < 0 \\ 0.8 - 0.6x & 0 \leq x \leq 1 \\ 0.2 & 1 < x \end{cases} \tag{3}$$

Clearly, 15 trigger points are insufficient to yield a smooth profile, and we typically use sets of several hundred trigger points. This could, in principle, lead to more than one monolayer of material being deposited per pass. In order to avoid this situation, the complete set of trigger points $x_n$ is first calculated, but only a subset is used for each pass:

for example, in pass one, the laser can be fired for n = 0, 5, 10, etc., in pass two for n = 1, 6, 11, etc.

Experimental results in Fig. 3 demonstrate the validity of this approach: The repeated deposition using a rate profile similar to that used in the simulations (but with 450 trigger points) yields a smooth variation, and repeated deposition of two profiles with opposite gradients results in a uniform thickness across the entire substrate area (45 mm). (For these experiments, $SrTiO_3$ was deposited at 780°C onto $Al_2O_3$ substrates, and the total thickness was about 300 nm for the uniform profile.)

As in all CCS approaches, the composition *c(x)* at each point *x* varies uniformly along the direction of the spread. Therefore, any sample of finite size $\Delta x$ bears a compositional variation of $\Delta c = (\partial c/\partial x) \Delta x$. The composition slope $(\partial c/\partial x)$ depends both on the range of compositions investigated ($\Delta c^{max}$) and the linear size of the spread ($\Delta x^{max}$), so that the composition non-uniformity of each sample is given by

$$\Delta c = \frac{\Delta c^{max}}{\Delta x^{max}} \Delta x . \qquad (4)$$

For example, in the case of an alloy $A_xB_{1-x}$, $\Delta c^{max} = 1$ if the entire range $0 \leq x \leq 1$ is investigated, but $\Delta c^{max} = 0.1$ if only alloys between $A_{0.3}B_{0.7}$ and $A_{0.4}B_{0.6}$ are of interest. For the examples of $(Sr_{1-x}Ca_x)RuO_3$ composition spreads given below, $\Delta c^{max} = 1$ and $\Delta x^{max} = 27$ mm, thus for a 2 mm wide sample, $\Delta c = 7.5\%$. This is sufficient for initial studies and the observation of general trends but can complicate detailed investigations. Therefore, such studies require a narrower concentration range (e.g. $\Delta c^{max} = 0.1$) chosen in the area of interest. Using our apparatus' capability of $\Delta x^{max} \geq 40$mm, a value of $\Delta c \leq 0.005$ can then readily be obtained.

Finally, it is possible to use the apparatus as described to grow (by simple two-target mixing and keeping the substrate stationary) uniform-composition films of any of the materials within a compositional spread, allowing for direct confirmation of the results and further detailed examination of samples with specific compositions.

A future generalization of this approach to (pseudo-)ternary phase diagrams will be straight-forward: it requires (1) a rotation of the substrate by 60° between the deposition steps of the individual constituents and (2) laser beam scanning along a line parallel to the aperture slit (to assure good deposition uniformity in that direction).

**$(Sr_{1-x}Ca_x)RuO_3$ composition spreads**

Binary composition-spread films of $(Sr_{1-x}Ca_x)RuO_3$ were grown according to the procedure described above, with a total film thickness of about 250 nm. Commercially available $SrRuO_3$ and $CaRuO_3$ targets were used for this study. 8 individual substrates of $LaAlO_3$ (each measuring 5 x 10 mm$^2$) were mounted on the heater plate within an area of 45 x 10 mm$^2$ (i.e. leaving less than 1 mm space between adjacent pieces). For detailed characterization, these samples were cut into 2 x 10 mm$^2$ slabs after deposition. Figure 4 shows measurements of the composition $c^{meas}(x)$ as function of position $x$ on the heater plate. The solid line indicates the profile $c^{design}(x)$ as entered into the control software. Measurements were performed using energy dispersive x-ray spectroscopy (EDX) in a scanning electron microscope, and by Rutherford Backscattering Spectroscopy (RBS); these two methods show agreement well within the expected experimental errors.

As is clearly seen, the experimental data agree well with the desired values for the composition. Two types of systematic deviations are nevertheless observed: First, there is clearly some "rounding" in the profile near the inflection points at 9 mm and 36 mm. This is expected for an aperture consisting of a slit with finite width. Furthermore, the experimental data points are systematically shifted by about 1 mm due to imperfect alignment of the system (laser spot, center of the aperture, and "zero position" on the heater position not falling onto a perfectly straight line). Taking these two effects into consideration, the expected composition variation $c^{calc}(x)$ can be calculated (by a superposition of Gaussian profiles centered at the "trigger points") and is shown as broken line in Fig. 4.

Figure 5 represents the same data in a different format. In Fig. 5a, the measured values of the composition are shown as function of the "design" parameter, i.e. $c^{meas}(c^{design})$. In Fig. 5b, the same data points are shown as function of the calculated values (i.e. $c^{meas}(c^{calc})$). Good linearity ($r > 0.998$ for $c^{meas}(c^{design})$) and good overall agreement is observed in both cases; even when plotted directly as a function of the desired compositions, deviations are mainly visible near the end-points. This illustrates that complicated calculations to obtain the composition values are not required, except perhaps in cases where a much wider mask is required to increase deposition speed. Even in such special cases the required calculations are simple and straight-forward.

Using the average squared difference between the measured composition and the design variable (i.e. $\sigma = N^{-1} \{ \Sigma (c^{meas} - c^{design})^2 \}^{-2}$) as a measure of the error in the prediction of the composition for each point of the spread, we find from data in Figs. 4 and 5 that the prediction of the composition is accurate to within less than 3%. The above-mentioned

"rounding" near the end-points, the misalignment of the system, and the error in positioning the tool for compositional analysis are likely to be the most significant contributions to this estimate. Thus the fundamental accuracy of the approach is expected to be significantly better.

**Properties of $(Sr_{1-x}Ca_x)RuO_3$ composition-spread films**

The $(Sr_{1-x}Ca_x)RuO_3$ films on $LaAlO_3$ substrates (cut into pieces measuring 2 x 10 mm$^2$) were first characterized using standard x-ray diffraction to determine their out-of-plane lattice parameters. The results are shown in Fig. 6, together with published data for polycrystalline bulk samples.[16] An expansion in the out-of-plane direction is observed, as would be expected for films grown under compressive strain on a substrate with a smaller lattice parameter. Excellent agreement is observed between the film and bulk data for the general trend of decreasing lattice parameter with increasing Ca concentration.

The sample's electrical resistance was measured as a function of temperature, and selected results are shown in Fig. 7. The resistivity at 300 K was 330 µΩcm for $SrRuO_3$, which compares favorably to literature values.[17,18] Finally, SQUID magnetometer measurements were performed, and preliminary results are shown in Fig. 8 for a sample with average composition of $Sr_{0.80}Ca_{0.20}RuO_3$ at 5 K. Despite the fact that $\Delta c = 0.08$ for this 2 x 10 mm$^2$ sample, the data (obtained with the magnetic field in the plane of the substrate) is similar to earlier reports for $SrRuO_3$,[19,20] again indicating that our approach yields high-quality samples. A more detailed investigation of the magnetic and transport properties of these samples is in progress.

**Conclusions**

We have introduced a CCS approach in which *in-situ* alloying occurs via the repeated, sequential deposition of sub-monolayer amounts of each constituents, using a PLD configuration. Contrary to our previous PLD-CCS method, the current technique yields samples with uniform thickness, and all portions of the sample are grown using the same deposition energetics. In addition, the advantages of the earlier technique, namely the applicability to meta-stable alloys and heterostructures, and the additional capability of the apparatus to form uniform films under identical conditions, are maintained. The current approach is based on the motion of the substrate rather than of a shutter, which makes deposition on larger areas possible without suffering from non-uniformities in deposition rates and energetics. This is of particular importance if samples are to be characterized by standard techniques, including temperature-dependent measurements of resistivity and dielectric and magnetic and properties (i.e. SQUID magnetometry).

The data presented in this work demonstrate that the composition across the substrate area varies precisely according to the desired profile, and that the obtained films are of high quality (as determined by a comparison between literature values and our data for $(Sr_{1-x}Ca_x)RuO_3$ composition-spread films). "Zooming-in" is easily accomplished in this approach, so that any portion of interest within phase spread can be analyzed in great detail.

**Acknowledgements**

**References**

[1] X.-D. Xiang, X. Sun, G. Briceno, Y. Lou, K. Wang, H. Chang, W.G. Wallace-Freedman, S. Chen, and P.G. Schultz, Science **268**, 1738 (1995).

[2] E. Danielson, J.H. Golden, E.W. McFarland, C.M. Reaves, W.H. Weinberg, and X.D. Wu, Nature **389**, 944 (1997).

[3] Y.K. Yoo, F. Duewer, H. Yang, D. Yi, J.-W. Li, and X.-D. Xiang, Nature **406**, 704 (2000).

[4] J. Li, F. Duewer, Ch. Gao, H. Chang, X.-D. Xiang, and Y. Lu, Appl. Phys. Lett. **76**, 769 (2000).

[5] K. Kennedy, *Atomic Energy Commission Report UCRL-16393*, Sept. 1965.

[6] J.J. Hanak, J. Mat. Sci. **5**, 964 (1970)

[7] R.B. van Dover, L.F. Schneemeyer, and R.M. Fleming, Nature **392**, 162 (1998).

[8] J.D. Perkins, J.A. del Cueto, J.L. Alleman, C. Warmsingh, B.M. Keyes, L.M. Gedvilas, P.A. Parilla, B. To, D.W. Readey, and D.S. Ginley, Thin Solid Films **411**, 152 (2002).

[9] P.K. Schenck and D.L. Kaiser, Proc. Knowledge Foundation, COMBI 2002, in press.

[10] H.-M. Christen, D.P. Norton, L.A. Géa, and L.A. Boatner, Thin Solid Films **312**, 156 (1998).

[11] H.M. Christen, S.D. Silliman, and K.S. Harshavardhan, Rev. Sci Instrum. **72**, 2673 (2001).

[12] H.M. Christen, S.D. Silliman, and K.S. Harshavardhan, Appl. Surf. Sci. **189**, 216 (2002).

[13] T. Fukumura, M. Ohtani, M. Kawasaki, Y. Okimoto, T. Kageyama, T. Koida, T. Hasegawa, Y. Tokura, and H. Koinuma, Appl. Phys. Lett. **77**, 3426 (2000).

[14] N. Okazaki, H. Odagawa, Y. Cho, T. Nagamura, D. Komiyama, T. Koida, H. Minami, P. Ahmet, T. Fukumura, Y. Matsumoto, M. Kawasaki, T. Chikyow, H. Koinuma, and T. Hasegawa, Appl. Surf. Sci. **189**, 222 (2002).

[15] C.N.R. Rao and B. Raveau, *Transition Metal Oxides*, 2nd ed. (John Wiley & Sons, New York, 1998).

[16] H. Kobayashi, M. Nagata, R. Kanno, and Y. Kawamoto, Mater. Res. Bull. **29**, 1271 (1994).

[17] C.B. Eom, R.J. Cava, R.M. Fleming, J.M. Phillips, R.B. van Dover, J.H. Marshall, J.W.P. Hsu, J.J. Krajewski, and W.F. Peck, Jr., Science **258**, 1766 (1992).

[18] L. Miéville, T.H. Geballe, L. Antognazza, and K. Char, Appl. Phys. Lett. **70**, 126 (1996).

[19] M. Izumi, K. Nakazawa, Y. Bando, Y. Yoneda, and H. Terauchi, J. Phys. Soc. Jpn. **66**, 3893 (1997)

[20] Q. Gan, R.A. Rao, C.B. Eom, J.L. Garrett, and M. Lee, Appl. Phys. Lett. **72**, 978 (1998).

[21] X.D. Wu, S.R. Foltyn, R.C. Dye, Y. Coulter, and R.E. Muenchausen, Appl. Phys Lett. **62**, 2434 (1993).

**Figure Captions**

**Figure 1.** Schematic representation of the PLD-CCS process. The substrate is passed behind a slit-shaped aperture, and the laser is fired whenever the substrate position coincides with one of the pre-determined "trigger points". Different sets of trigger points are used for the two constituents of the alloy, and less than one monolayer of material is deposited in each cycle.

**Figure 2.** Calculated results for deposition rate or thickness profiles assuming a Gaussian distribution of the deposited material with a width of 0.1 for each laser pulse. The dotted line in (a) indicates the desired profile. Clearly, the use of 15 trigger points is insufficient to obtain a uniform profile.

**Figure 3.** Thickness profiles (measured by profilometry) of $SrTiO_3$ films grown on $Al_2O_3$. Circles represent the thickness of a single "wedge" profile obtained by repeatedly passing the substrate behind the aperture and using a set of 450 trigger points. Crosses indicate the thickness profile obtained by the overlapping of two "wedges" with opposite gradient. Error bars correspond to the uncertainty of the profilometry measurements.

**Figure 4.** Ca-concentration as measured on 16 points along a composition spread of $(Sr_{1-x}Ca_x)RuO_3$, measured by EDX (circles) and RBS (crosses). The desired profile (i.e. the parameters supplied to the control software) is indicated by the solid line. The broken

line is the calculated composition assuming an aperture with a finite slit width, and a 1-mm misalignment between laser spot, slit center, and heater position.

**Figure 5.** (a) Measured Ca concentration for the same composition spread as in Fig. 4, but plotted as function of the desired Ca content. In (b), the same data is drawn as function of the calculated Ca content (i.e. the broken line in Fig. 4). In both cases, the linearity is excellent, and differences between (a) and (b) are observed mainly near the end-points of the spread. This shows that complicated calculations to obtain the composition as function of position on the substrate are not required for most applications.

**Figure 6.** Lattice parameter as function of Ca-concentration for $(Sr_{1-x}Ca_x)RuO_3$ composition-spread films on $LaAlO_3$ substrates. The values obtained for these films are slightly larger than those for bulk polycrystalline samples, as expected for films grown under in-plane compressive strain.
[1]Data from Ref. 16.

**Figure 7.** Resistance as function of temperature for 3 representative samples of the $(Sr_{1-x}Ca_x)RuO_3$ composition spread, normalized at T = 300K. Arrows indicate approximate positions of the ferromagnetic transition. Values of the resistance at 300 K are below 350 µΩcm for all samples. Circles represent published data for a sputtered film after lift-off from the $SrTiO_3$ substrate,[20] squares for a PLD-grown film on $LaAlO_3$.[21]

**Figure 8.** Magnetic hysteresis loop at 5K for a $(Sr_{1-x}Ca_x)RuO_3$ sample across which x varies from 0.16 to 0.24. Sample size is 2 x 10 mm$^2$, and this preliminary data shows that our method yields samples large enough for such experiments.

**Figure 1.** Schematic representation of the PLD-CCS process. The substrate is passed behind a slit-shaped aperture, and the laser is fired whenever the substrate position coincides with one of the pre-determined "trigger points". Different sets of trigger points are used for the two constituents of the alloy, and less than one monolayer of material is deposited in each cycle.

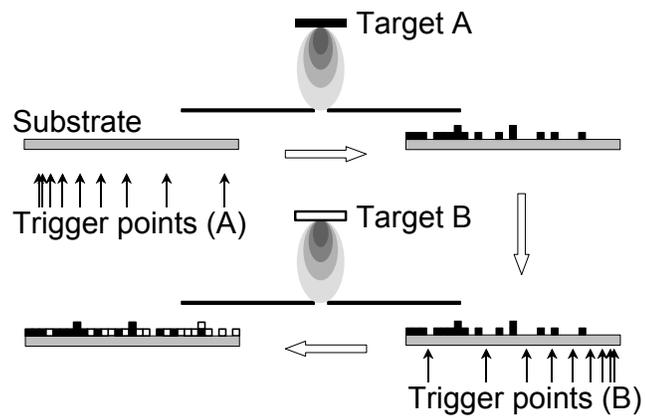

Figure 1, H.M. Christen *et al.*, Rev. Sci. Instrum.

**Figure 2.** Calculated results for deposition rate or thickness profiles assuming a Gaussian distribution of the deposited material with a width of 0.1 for each laser pulse. The dotted line in (a) indicates the desired profile. Clearly, the use of 15 trigger points is insufficient to obtain a uniform profile.

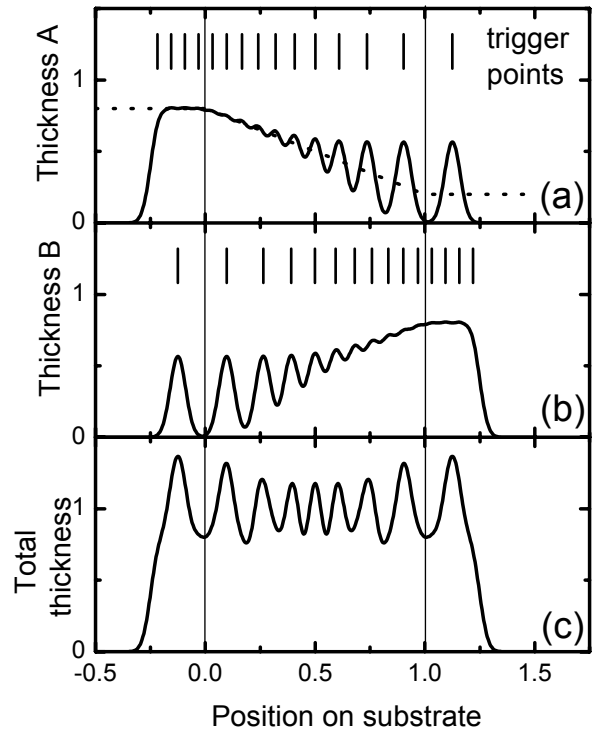

Figure 2, H.M. Christen *et al.*, Rev. Sci. Instrum.

**Figure 3.** Thickness profiles (measured by profilometry) of $SrTiO_3$ films grown on $Al_2O_3$. Circles represent the thickness of a single "wedge" profile obtained by repeatedly passing the substrate behind the aperture and using a set of 450 trigger points. Crosses indicate the thickness profile obtained by the overlapping of two "wedges" with opposite gradient. Error bars correspond to the uncertainty of the profilometry measurements.

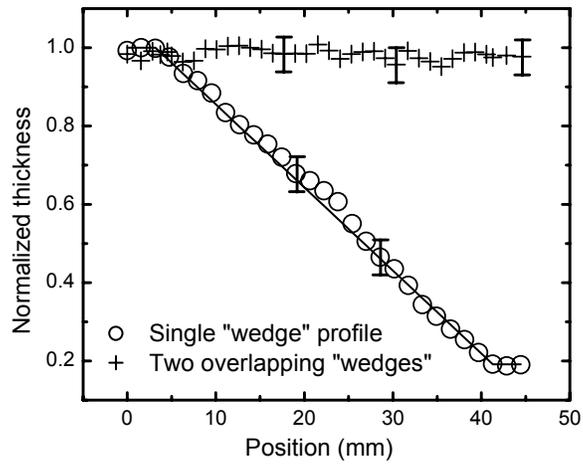

Figure 3, H.M. Christen *et al.*, Rev. Sci. Instrum.

**Figure 4.** Ca-concentration as measured on 16 points along a composition spread of $(Sr_{1-x}Ca_x)RuO_3$, measured by EDX (circles) and RBS (crosses). The desired profile (i.e. the parameters supplied to the control software) is indicated by the solid line. The broken line is the calculated composition assuming an aperture with a finite slit width, and a 1-mm misalignment between laser spot, slit center, and heater position.

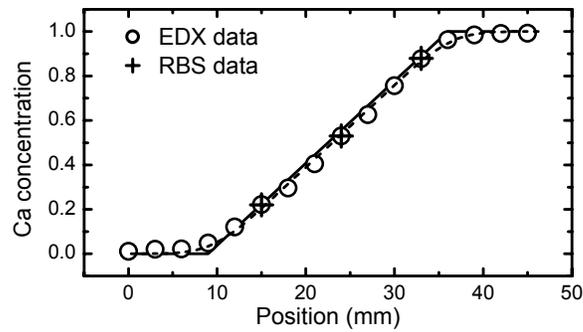

Figure 4, H.M. Christen *et al.*, Rev. Sci. Instrum.

**Figure 5.** (a) Measured Ca concentration for the same composition spread as in Fig. 4, but plotted as function of the desired Ca content. In (b), the same data is drawn as function of the calculated Ca content (i.e. the broken line in Fig. 4). In both cases, the linearity is excellent, and differences between (a) and (b) are observed mainly near the end-points of the spread. This shows that complicated calculations to obtain the composition as function of position on the substrate are not required for most applications.

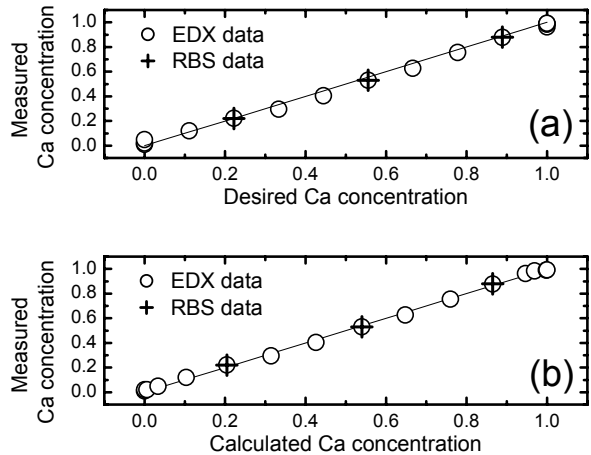

Figure 5, H.M. Christen *et al.*, Rev. Sci. Instrum.

**Figure 6.** Lattice parameter as function of Ca-concentration for $(Sr_{1-x}Ca_x)RuO_3$ composition-spread films on $LaAlO_3$ substrates. The values obtained for these films are slightly larger than those for bulk polycrystalline samples, as expected for films grown under in-plane compressive strain.
[1]Data from Ref. 16.

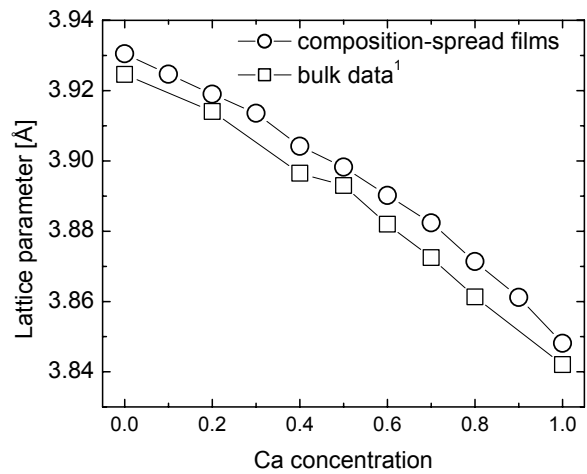

Figure 6, H.M. Christen *et al.*, Rev. Sci. Instrum.

**Figure 7.** Resistance as function of temperature for 3 representative samples of the $(Sr_{1-x}Ca_x)RuO_3$ composition spread, normalized at T = 300K. Arrows indicate approximate positions of the ferromagnetic transition. Values of the resistance at 300 K are below 350 μΩcm for all samples. Circles represent published data for a sputtered film after lift-off from the $SrTiO_3$ substrate,[20] squares for a PLD-grown film on $LaAlO_3$.[21]

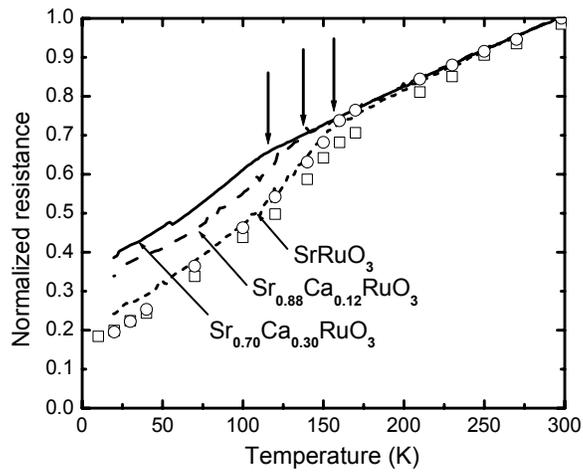

Figure 7, H.M. Christen *et al.*, Rev. Sci. Instrum.

**Figure 8.** Magnetic hysteresis loop at 5K for a $(Sr_{1-x}Ca_x)RuO_3$ sample across which x varies from 0.16 to 0.24. Sample size is 2 x 10 mm$^2$, and this preliminary data shows that our method yields samples large enough for such experiments.

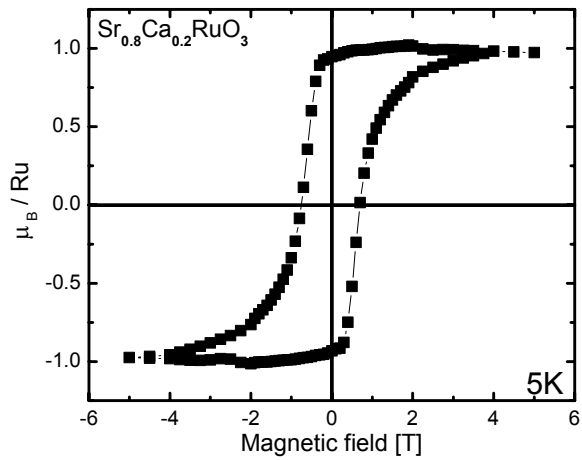

Figure 8, H.M. Christen *et al.*, Rev. Sci. Instrum.